\documentclass[prb,twocolumn,floatfix,showpacs,preprintnumbers,amsmath,amssymb]{revtex4}

\usepackage{graphicx}
\usepackage{dcolumn}
\usepackage{bm}

\begin{document}

\preprint{APS/123-QED}

\title{Successive phase transitions to antiferromagnetic and weak-ferromagnetic long-range orders in quasi-one-dimensional antiferromagnet Cu$_3$Mo$_2$O$_9$
}

\author{Tomoaki Hamasaki}
 \email{t-hamasa@sophia.ac.jp}
\author{Tomoyuki Ide}
\author{Haruhiko Kuroe}
\author{Tomoyuki Sekine}
\affiliation{%
Department of Physics, Sophia University, 7-1 Kioi-cho, Chiyoda, Tokyo 102-8554, Japan
}%

\author{Masashi Hase}
\affiliation{%
National Institute for Materials Science (NIMS), 1-2-1 Sengen, Tsukuba 305-0047, Japan
}%

\author{Ichiro Tsukada}
\affiliation{%
Central Research Institute of Electric Power Industry, 2-11-1 Iwadokita, Komae-shi, Tokyo 201-8511, Japan
}%

\author{Toshiro Sakakibara}
\affiliation{%
Institute for Solid State Physics, University of Tokyo, 5-1-5 Kashiwanoha, Kashiwa-shi, Chiba 277-8581, Japan
}%

\date{\today}

\begin{abstract}

Investigation of the magnetism of Cu$_3$Mo$_2$O$_9$ single crystal, which has antiferromagnetic (AF) linear chains interacting with AF dimers, reveals an AF second-order phase transition at $T_{\rm N} = 7.9$ K.
Although weak ferromagnetic-like behavior appears at lower temperatures in low magnetic fields, complete remanent magnetization cannot be detected down to 0.5 K. 
However, a jump is observed in the magnetization below weak ferromagnetic (WF) phase transition at $T_{\rm c} \simeq 2.5$ K when a tiny magnetic field along the $a$ axis is reversed, suggesting that the coercive force is very weak.
A component of magnetic moment parallel to the chain forms AF long-range order (LRO) below $T_{\rm N}$, while a perpendicular component is disordered above $T_{\rm c}$ at zero magnetic field and forms WF-LRO below $T_{\rm c}$.
Moreover, the WF-LRO is also realized with applying magnetic fields even between $T_{\rm c}$ and $T_{\rm N}$.
These results are explainable by both magnetic frustration among symmetric exchange interactions and competition between symmetric and asymmetric Dzyaloshinskii-Moriya exchange interactions.

\end{abstract}

\maketitle

\section{Introduction}

The magnetic frustration among symmetric exchange interactions 
in some magnetic materials can be so intense that it induces novel and complex phenomena.
Examples are a spin ice state in three-dimensional pyrochlore-lattice antiferromagnetic (AF) systems \cite{Ramirez1999,Bramwell2001} and a spin nematic phase in triangular antiferromagnets.\cite{Nakatsuji2005}
The competition between symmetric and asymmetric Dzyaloshinskii-Moriya (DM) exchange interactions creates diverse types of magnetic long-range order such as weak ferromagnetic long-range order (WF-LRO).
Intriguing magnetic properties should arise in spin systems with both frustration and competition.
In Ni$_{3}$V$_{2}$O$_{8}$, successive phase transitions occur with decreasing temperature, and two incommensurate and two commensurate phases appear.
WF-LRO in the commensurate phase and ferroelectricity in the incommensurate phase have been observed.\cite{Lawes2004,Lawes2005}
CuB$_{2}$O$_{4}$ undergoes successive phase transitions to commensurate and incommensurate phases with decreasing temperature.
WF-LRO is found in the commensurate phase \cite{Petrakovskii1999,Petrakovskii2000} and a soliton lattice appears at around the temperature of transition to the incommensurate phase without applied magnetic fields.\cite{Roessli2001}\par
We investigate a quasi-one dimensional spin system Cu$_3$Mo$_2$O$_9$, because it should have both magnetic frustration among symmetric exchange interactions and competition between symmetric and asymmetric exchange interactions in addition to the one-dimensional fluctuation. 
In this paper, we report our findings of the coexistence of antiferromagnetic long-range order (AF-LRO) and spin disorder at a single site, and of the successive phase transition to WF-LRO.

\par

\section{Crystal Structure and spin system in C\lowercase{u}$_3$M\lowercase{o}$_2$O$_9$}

\begin{figure}[t]
	\begin{center}
\includegraphics[width=.45\textwidth]{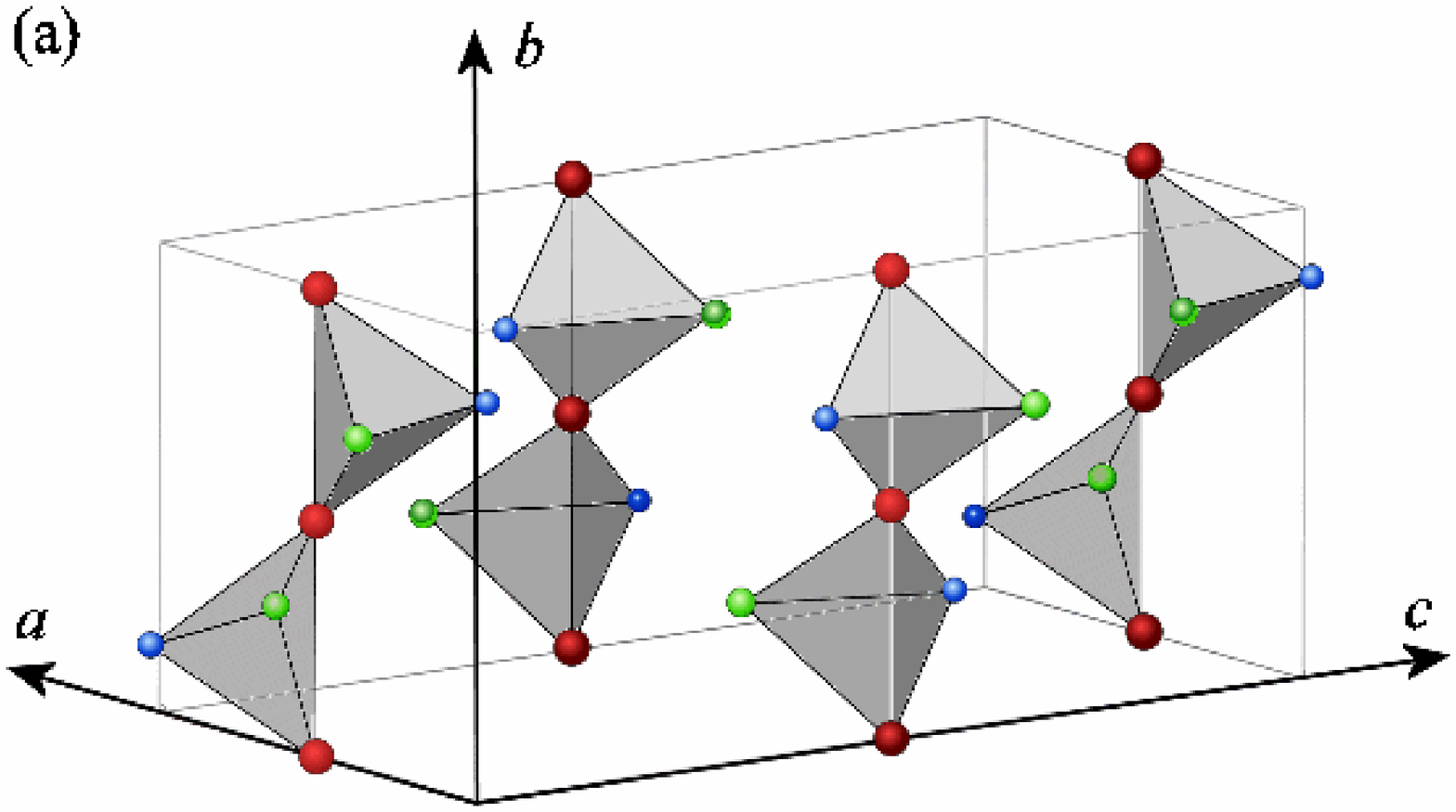}
\includegraphics[width=.45\textwidth]{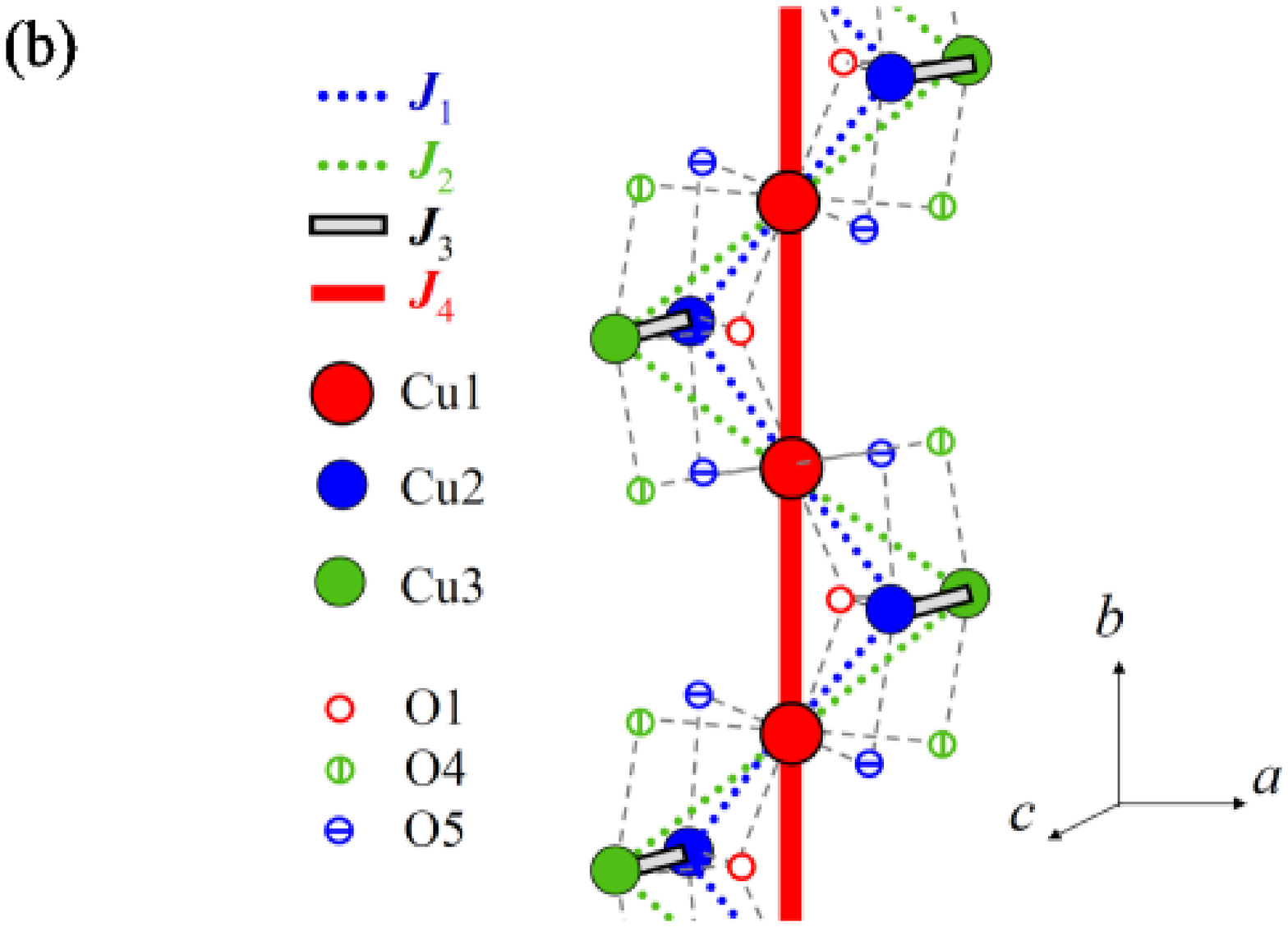}
\caption{(color online) (a) Schematic drawing of Cu$^{2+}$ ion position in Cu$_3$Mo$_2$O$_9$ and (b) structure of the tetrahedral spin chain with the Cu-O-Cu paths.
}
  	\label{fig:cmopos01}
	\end{center}
  \end{figure}

The space group of Cu$_3$Mo$_2$O$_9$ is orthorhombic $Pnma$ 
and the lattice parameters are $a = 7.6793$ \AA, $b = 6.8728$ \AA, 
and $c = 14.6222$ \AA \ at room temperature.\cite{Reichelt1997,Reichelt2005}
Only the Cu$^{2+}$ ions have spin-1/2, while other ions are nonmagnetic.
There are three crystallographically inequivalent Cu$^{2+}$ sites (Cu1, Cu2, and Cu3).
As shown in Fig. \ref{fig:cmopos01}, there are four Cu1 sites in a unit cell, i.e., $Z = 4$.
\par
\begin{table*}[t]
\caption{\label{tab:1}
Interatomic distances and angles of Cu$_3$Mo$_2$O$_9$.
The suffix of $J$ was sorted by short Cu-Cu distance. 
$J_1$ and $J_2$ have two Cu-O-Cu paths and angles as 
shown in Fig. \ref{fig:cmopos01}(b).}
\begin{ruledtabular}
\begin{tabular}{ccccc}
& Bond 1 & Bond 2 & Bond 3 & Bond 4 \\
& (Cu1-Cu2) & (Cu1-Cu3) & (Cu2-Cu3) & (Cu1-Cu1) \\
\hline
Interaction&$J_1$&$J_2$&$J_3$&$J_4$\\
Cu-Cu distance & 2.95 \AA & 3.00 \AA & 3.17 \AA & 3.44 \AA \\
\begin{tabular}{c}
Cu-O-Cu path \\
Cu-O-Cu angle (F or AF)
\end{tabular}
& 
\begin{tabular}{rl}
Cu1-O1-Cu2, & Cu1-O5-Cu2 \\
101.9$^\circ$ (AF), & 88.2$^\circ$ (F) 
\end{tabular}
&
\begin{tabular}{rl}
Cu1-O1-Cu3, & Cu1-O4-Cu3 \\
103.8$^\circ$ (AF), & 91.0$^\circ$ (F)
\end{tabular}   
&
\begin{tabular}{c}
Cu2-O1-Cu3 \\
109.2$^\circ$ (AF)
\end{tabular}
&
\begin{tabular}{c}
Cu1-O1-Cu1 \\
134.9$^\circ$ (AF)
\end{tabular}
\end{tabular}
\end{ruledtabular}
\end{table*}
We first describe the expected magnetic interactions. 
As shown in Table {\ref{tab:1}}, there are four kinds of short Cu-Cu bonds (Bonds 1 to 4),
and the corresponding exchange parameters are defined as $J_i$ in bond $i$.
The signs and magnitudes of exchange interactions are determined mainly by Cu-O-Cu angles and Cu-Cu distances.
It is reasonable to expect that the $J_3$ and $J_4$ interactions are AF. 
Because the angle of bond 4 (Cu1-O1-Cu1 $= 134.9^{\circ}$) is larger than 
that in bond 3 (Cu2-O1-Cu3 $= 109.2^{\circ}$), the $J_4$ interaction is expected to be stronger than $J_3$.\cite{Mizuno1998} 
In bond 1, the interactions in the two Cu-O-Cu paths with bond angles of Cu1-O1-Cu2 $=101.9^{\circ}$ and Cu1-O5-Cu2 $=88.2^{\circ}$ are probably AF and ferromagnetic (F), respectively. 
These two interactions may cancel each other out. 
Thus, the sign of $J_1$ cannot be easily judged. 
It is inferred that the magnitude of the $J_1$ interaction is smaller than that of the $J_3$ and $J_4$ ones. 
A similar assumption is applicable to the $J_2$ interaction (See also Table {\ref{tab:1}}). 
The exchange interactions in the other bonds are much weaker than those for $J_1$ - $J_4$, 
because the other Cu-Cu distances are longer than 5.01 \AA.
Consequently, the $J_3$ and $J_4$ interactions are the main determinants of the magnetism of Cu$_3$Mo$_2$O$_9$. 
The $J_4$ interaction forms uniform AF linear chains of the Cu1 spins in the $b$ direction, and the $J_3$ interaction forms AF dimers of the Cu2 and Cu3 spins.
Every chain is coupled to neighboring dimers by the $J_1$ and $J_2$ interactions.
The unit cell consists of two tetrahedral spin chains, as seen in Fig. \ref{fig:cmopos01}.
We emphasize that the spin system in Cu$_3$Mo$_2$O$_9$ is different from the diamond chain, because the $J_4$ interaction does not exist in the diamond chain.\cite{Okamoto2003}

\section{Experiments}

Single crystals of Cu$_3$Mo$_2$O$_9$ were prepared a flux method 
of mixtures with Rb$_2$MoO$_4$, CuMoO$_4$ and MoO$_3$ so as 
to be Rb : Cu : Mo = 2 : 1 : 3.
A mixture of these materials was sintered at 700 $^\circ$C 
and cooled down to 490 $^\circ$C for a week.
We obtained millimeter-sized ($1 \times 5 \times 3$ mm$^3$) and slightly reddish black single crystals.
We identified the ground single crystal as Cu$_3$Mo$_2$O$_9$ 
by comparing the X-ray diffraction pattern with that of the powder sample.
Moreover, absence of Rb$^+$ was checked by X-ray photoemission spectroscopy.
The temperature dependence of magnetic susceptibility $M/H$ 
and magnetic-field dependence of magnetization $M$ above 2 K
were measured using SQUID magnetometers 
(Quantum Design MPMS-5S and Conductus $\chi$MAG).
The magnetization below 2 K was obtained 
using Faraday magnetometer with $^3$He refrigerator.\cite{Sakakibara1994}
The specific heat was measured at zero magnetic field 
by thermal relaxation method (Quantum Design PPMS).

\par

\section{EXPERIMENTAL RESULTS}

  \begin{figure}[h]
	\begin{center}
\includegraphics[width =.94\linewidth]{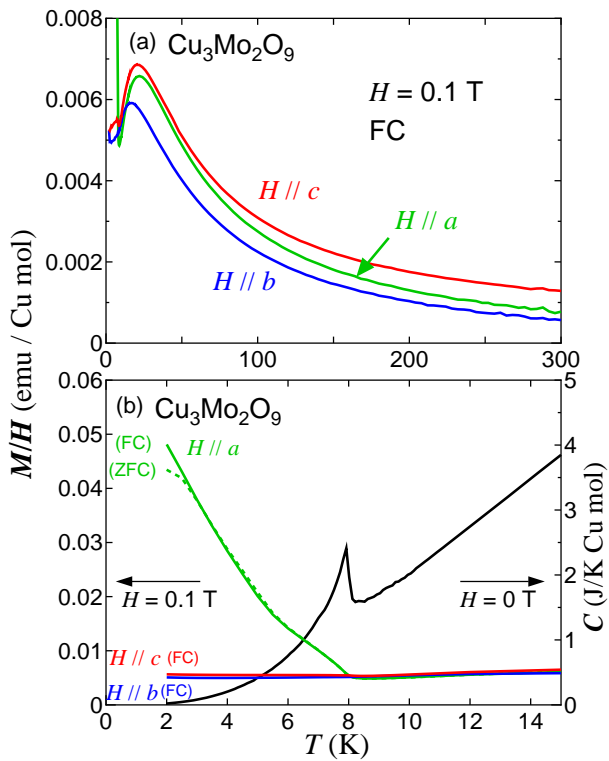}
\caption{(color online) 
(a) Temperature dependence of $M/H$ below 300 K for three crystal-axis directions in Cu$_3$Mo$_2$O$_9$. 
(b) Temperature dependence of $M/H$ and specific heat $C$ below 15 K.
ZFC and FC mean the zero-field-cooling and field-cooling processes, respectively.
}\label{fig:expMT}
	\end{center}
  \end{figure}

Figure \ref{fig:expMT}(a) shows the temperature dependence of $M/H$ of a Cu$_3$Mo$_2$O$_9$ single crystal in a magnetic field $H$ at 0.1 T. 
The magnetizations over $H$ parallel to the $a$, $b$, and $c$ axes ($M_{a}/H$, $M_{b}/H$, and $M_{c}/H$) 
show broad peaks at 23, 16, and 21 K, respectively.
A broad peak is characteristic of low-dimensional antiferromagnets and/or spin dimers.
In addition to the differences in peak positions, the overall temperature dependences in the three directions differed from one another.
These differences cannot be explained by only anisotropy of the temperature-independent $g$-factors. 
Figure \ref{fig:expMT}(b) shows the temperature dependence of $M/H$ and specific heat $C$ below 15 K.
We found a drastic increase in $M_{a}/H$ 
below 7.9 K with decreasing temperature. 
In addition, a sharp $\lambda$-type peak in $C$ was observed at 7.9 K.
These results clearly indicate that a second-order magnetic phase transition 
occurs at $T_{\rm N}=7.9$ K.
The value of $M_{a}$ at 2.0 K corresponds to 0.7\% of the perfectly saturated value of spin-1/2 magnetic system. 
This small value immediately eliminates a possibility of ferromagnetic long-range order. 
Ferrimagnetic long-range order is also impossible because both the $J_4$ and $J_3$ interactions are AF.
Consequently, the rapid increase in $M_a /H$ below $T_{\rm N}$ suggests an appearance of weak ferromagnetic long-range order (WF-LRO) at finite magnetic fields. 
However, as shown later, the WF-LRO is not stabilized at zero magnetic field.
Below $\sim$2.5 K, moreover, temperature hysteresis was clearly observed in $M_{a}/H$ between the zero-field-cooling (ZFC) and field-cooling (FC) processes. 
Small changes in $M_{b}/H$ and $M_{c}/H$ were also observed at $T_{\rm N}$, 
although the change is not apparent in the scale of the vertical axis in the figure.
Temperature hysteresis was not detected for $M_b /H$ and $M_c /H$. 
\par
  \begin{figure}[h]
	\begin{center}
\includegraphics[width =.94\linewidth]{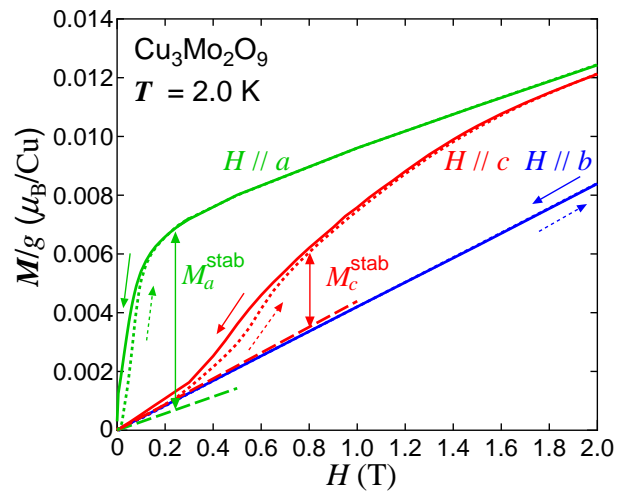}
\caption{(color online) 
Magnetization $M$ of Cu$_3$Mo$_2$O$_9$ at 2.0 K.
Green and red dashed lines represent terms linearly proportional to $H$.
Definitions of $M_{a}^{\rm stab}$ and $M_{c}^{\rm stab}$ are written in text.
}\label{fig:expMH}
	\end{center}
  \end{figure}
Figure \ref{fig:expMH} shows the $M$-$H$ curve at 2.0 K.
$M_{a}$ increased rapidly with the magnetic field.
Hysteresis appears below 0.2 T 
and it disappeared above 0.2 T.
$M_{c}$ increased slowly with the magnetic field below about 0.3 T.
At $0.3 < H < 0.8$ T, a rapid increase with finite hysteresis was observed.
With increasing magnetic field between 2.0 and 5.0 T, 
$M_{c}$ asymptotically approached $M_{a}$.
$M_{b}$ increased almost linearly without hysteresis below 5.0 T.
  \begin{figure}[h]
	\begin{center}
\includegraphics[scale =.45]{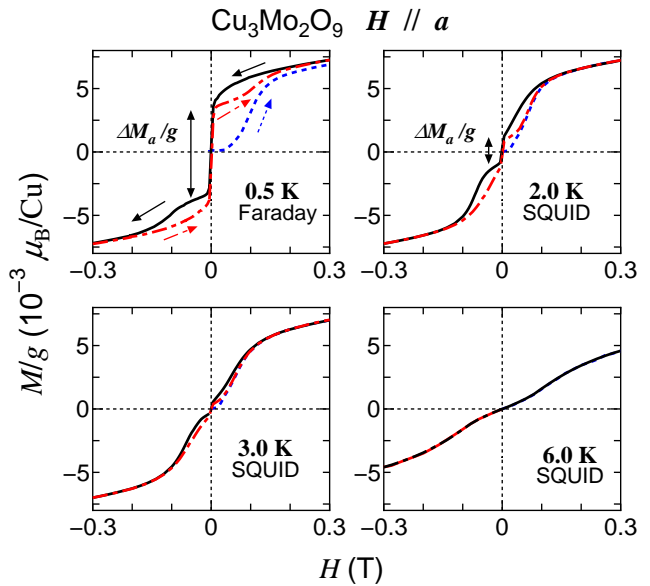}
\caption{(color online) 
Detail of $M_a$ of Cu$_3$Mo$_2$O$_9$ around $H = 0$ T 
at 0.5 K, 2.0 K, 3.0 K, and 6.0 K.
The blue dotted curves indicate the field-applying process after zero field cooling.
The black solid and red dash-dotted curves indicate the field-decreasing and 
field-applying processes, respectively.
}
  	\label{fig:MHdet}
	\end{center}
  \end{figure}
Because $M_a$ increased rapidly at low $H$, 
we measured the detailed behavior of $M_a$ every $\pm 50$ oersteds around $H = 0$ T
at 0.5, 2.0, 3.0, and 6.0 K, as shown in the Fig. \ref{fig:MHdet}.
The magnetic hysteresis did not appear below 0.3 T at 6.0 K, while the double 
hysteresis loops were clearly observed below 3.0 K.
However, the magnetization always crossed zero at $H=0$ T even below 3.0 K.
\par

The $M$-$H$ curve in close proximity to $H = 0$ T at 0.5 K 
is changed drastically with a tiny magnetic fields.
  \begin{figure}[h]
	\begin{center}
\includegraphics[scale =.45]{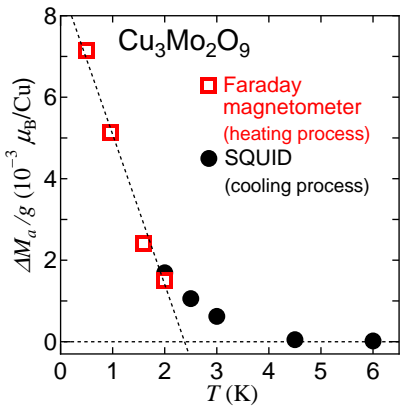}
\caption{(color online) 
Temperature dependence of the jumped magnetization $\Delta M_a/g$ at $H = 0$ T along the $a$ axis.
}
  	\label{fig:t_c}
	\end{center}
  \end{figure}
Applying only $+50$ Oe, $M_a$ reaches $+3.5 \times 10^{-3}$ $\mu_{\rm B}$/Cu and $M_a$ increased gradually above $+50$ Oe with magnetic hysteresis.
When the inverse field of $-50$ Oe is applied, $M_a$ goes to $-3.3 \times 10^{-3}$ $\mu_{\rm B}$/Cu.
The coercive force cannot be detected within a few tens Oe, taking into account the interval of $\pm 50$ Oe in the measurement of $M/H$ and the residual magnetic field of the superconducting magnet.
Figure \ref{fig:t_c} shows the jumped magnetization ($\Delta M_a/g$) as a function of temperature.
The jumped magnetization $\Delta M_a/g$ at 0 T was obtained by extrapolating from the values between $+50$ and $+400$ Oe and between $-50$ and $-400$ Oe.
It may be regarded as a spontaneous magnetization when the coercive force is very weak, and it vanished at about 3.5 K.
Taking into account the above-mentioned ambiguity of the applied magnetic field, we think that the WF phase transition occurs at $T_{\rm c} \simeq 2.5$ K at zero magnetic field in Cu$_{3}$Mo$_{2}$O$_{9}$. 
We also observed the temperature hysteresis in $M_{a}/H$ between the zero-field-cooling (ZFC) and field-cooling (FC) processes below $T_{\rm c} \simeq 2.5$ K.
We think that the component for the $a$ axis of magnetic moments is completely stabilized when the hysteresis in $M$-$H$ curve is closed at a magnetic field, e.g., at $H \sim 0.2$ T at 0.5 K, as seen in Fig. \ref{fig:MHdet}.
\par
No anomaly in the specific heat was observed at $T_{\rm c}$, as shown in Fig. \ref{fig:expMT}(b).
As discussed later, since this phase transition originates from the ordering of the component for the $a$ axis of magnetic moment with $\sim 10^{-2}$ $\mu_{\rm B}$/Cu, the specific-heat anomaly is expected to be very small at $T_{\rm c}$.
It is, therefore, difficult for us to detect it with specific heat measurement.
A rapid increase of $M_c$ with hysteresis was observed at 2.0 K when a magnetic field more than 0.3 T was applied along the $c$ axis, suggesting that a finite component of the WF-LRO parallel to the $c$ axis was induced by the applied magnetic field.
However, we did not observe transverse magnetization above 2.0 K in Cu$_{3}$Mo$_{2}$O$_{9}$.
Tsukada {\it et al}. reported that the AF phase transition took place together with the formation of WF-LRO at $T_{\rm N} = 8.8$ K in BaCu$_{2}$Ge$_{2}$O$_{7}$.\cite{Tsukada2000}
They observed the spontaneous magnetization along the $b$ axis.
Moreover, the weak magnetic field along the $a$ axis can change the direction of the magnetization to the $a$-axis direction, resulting in a disappearance of the transverse magnetization, i.e., the $b$-axis net magnetization.
The absence of the transverse magnetization indicates that the weak 
ferromagnetism of Cu$_3$Mo$_2$O$_9$ is more complicated than 
that of BaCu$_2$Ge$_2$O$_7$, which has only one equivalent Cu$^{2+}$ sites.
\par

\section{DISCUSSION}

\subsection{$J_3$ and $J_4$ interactions}

\begin{figure}[t]
	\begin{center}
\includegraphics[scale =.49]{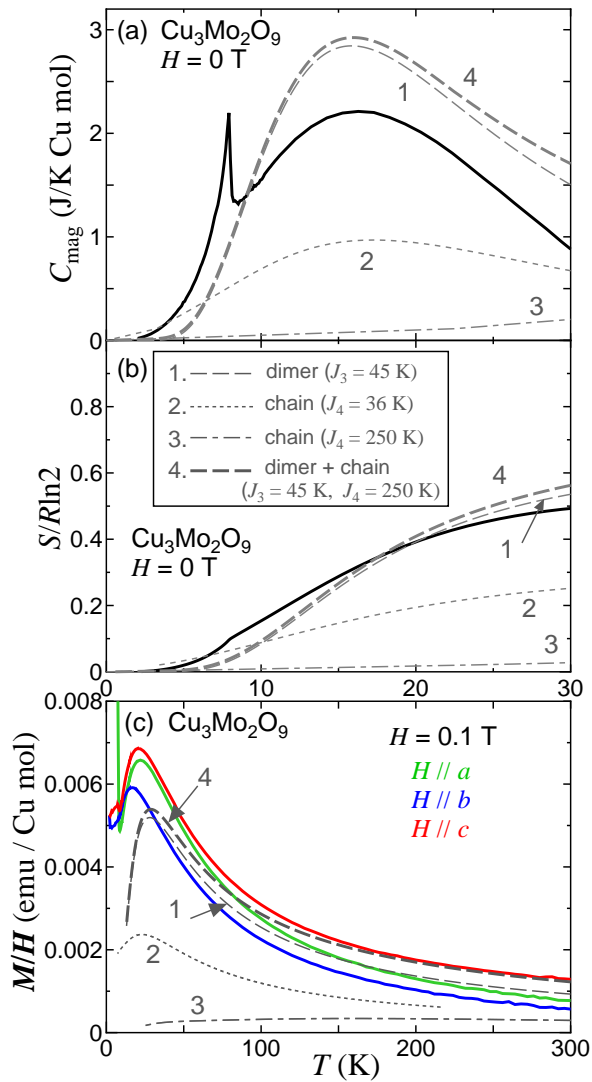}
\caption{(color online) 
Comparison (a) between the observed magnetic specific heat $C_{\rm mag}$ and 
calculated $C_{\rm mag}$, 
(b) between the observed magnetic entropy $S/R{\rm ln}2$ and 
calculated $S/R{\rm ln}2$, and 
(c) between the observed magnetic susceptibilities $M/H$ 
and the calculated susceptibilities in Cu$_3$Mo$_2$O$_9$.
The several thick solid curves denote the experimental results.
The gray and thin dashed curves (line 1), and 
the gray and thin dotted ones (line 2), and 
the gray and thin dash-dotted ones (line 3) denote 
the calculations for 
the isolated spin dimers ($J_3 = 45$ K), 
the uniform AF linear chains ($J_4 = 36$ K), and
the uniform AF linear chains ($J_4 = 250$ K), respectively.
The gray and thick dashed curves (line 4) denote 
the calculation in the case of $J_{3} = 45$ K and $J_{4} = 250$ K in Cu$_3$Mo$_2$O$_9$.
}\label{fig:fit}
	\end{center}
  \end{figure}

First we roughly estimate the magnitudes of the $J_3$ and $J_4$ interactions, which are expected to play dominant roles in the magnetism of Cu$_{3}$Mo$_{2}$O$_{9}$, as written in Section II.
As a zeroth approximation, we compared the magnetic specific heat $C_{\rm mag}$, the magnetic entropy $S/R{\rm ln}2$, and the magnetic susceptibility $M/H$ with those of the noninteracting AF linear chains and isolated spin dimers in Fig. \ref{fig:fit}.
Here, $R$ is the gas constant.
$C_{\rm mag}$ was obtained by subtracting 
the specific heat of the nonmagnetic and almost 
isomorphic Zn$_3$Mo$_2$O$_9$ powder, 
which is expected to have almost the same phonon dispersions as Cu$_3$Mo$_2$O$_9$, because the mass of Zn approximately equals that of Cu.

As shown in Fig. \ref{fig:fit}(a), the magnetic specific heat $C_{\rm mag}$ shows a broad hump peaked at about 16 K. 
This behavior is characteristic of one-dimensional antiferromagnets or isolated spin dimers.
We also calculated magnetic entropy. 
With increasing temperature, the entropy rapidly increases and then the increase was moderated.
The value of magnetic entropy at 30 K is much smaller than the high-temperature limit of entropy ($R{\rm ln}2$), as shown in the Fig. \ref{fig:fit}(b).
Taking into account the temperature dependence of the entropy in the AF linear chain and that in the isolated dimer in Fig. 4 of ref. \onlinecite{BF1969}, we conclude that the entropy in the AF linear chain or that in the isolated dimer should be released at high temperatures far above 30 K.
It means that there exists a strong AF interaction of which magnitude is in the order of hundreds Kelvin.
We discuss whether this peak originates from the spin dimer of Cu2 and Cu3 ($J_3$) or the spin chain of Cu1 ($J_4$) by calculating $C_{\rm mag}$.
These calculated specific heats are expressed as Schottky-type one and Bonner-Fisher one,\cite{BF1969} respectively.
Judging from the peak temperature of $C_{\rm mag}$, as shown in gray and thin curves in Fig. \ref{fig:fit}(a), either parameter of $J_{3} = 45$ K (line 1) or $J_{4} = 36$ K (line 2) is possible.
However, the absolute value of calculated $C_{\rm mag}$ with $J_{4} = 36$ K is much smaller than that of experimental one.

Figure \ref{fig:fit}(c) compares $M/H$s along the $a$, $b$, and $c$ axes to the calculated one with $J_3 = 45$ K (line 1) and that with $J_4 = 36$ K (line 2).
We used the Bonner-Fisher curve for the spin chain of Cu1 ($J_4$) and the calculation for the isolated spin dimer of Cu2 and Cu3 ($J_3$).
In the calculation, we set the $g$-factor to be 2.154 which was the root mean square of [$(g_{a}, g_{b}, g_{c}) = (2.090, 2.193, 2.180)$] obtained by ESR measurements using an $X$-band spectrometer at room temperature.
The ambiguity of calculated susceptibility owing to the $g$-factor was less than 10\%.
The temperature-independent susceptibility generated by the Van-Vleck paramagnetism is negligible because its magnitude, roughly depending on the species of ions, is in the order of $10^{-4}$ emu/Cu mol.
The calculated susceptibility of the uniform AF linear chain with $J_{4} = 36$ K, as shown by line 2 in Fig. \ref{fig:fit}(c), cannot reproduce the experimental data, while that of the spin dimer with $J_{3} = 45$ K can roughly reproduce the experimental one.
Therefore, it is understood that the peaks in $C_{\rm mag}$ and $M/H$ come mainly from the magnetic properties of the spin dimers.
We studied inelastic neutron scattering of Cu$_3$Mo$_2$O$_9$ powders and observed a broad peak of magnetic excitation at about 4 meV ($\sim 46$ K).
This peak probably originates from the spin gap of the spin dimer between the spin-singlet and triplet states,\cite{Hamasaki_neutron} supporting the present estimation.

Next, we estimated the parameter of $J_4$ interaction by comparing $M/H$ with the calculated susceptibility, in particular in the high-temperature region.
As stated above, we estimated that the $J_4$ interaction is much stronger than the $J_3$ one.
It is consistent with the relation between exchange interactions and Cu-O-Cu bonds, because the angle of bond 4 (Cu1-O1-Cu1 $=134.9^{\circ}$) is larger than 
that in bond 3 (Cu2-O1-Cu3 $=109.2^{\circ}$),\cite{Mizuno1998} as shown in Table {\ref{tab:1}}.  
One of the rough estimations is given by the case of $J_{4} = 250$ K, which is denoted by line 3 in Fig. \ref{fig:fit}.
The summation of the calculated susceptibilities with $J_{3} = 45$ K and $J_{4} = 250$ K are shown by line 4 in Fig. \ref{fig:fit}.
This calculated result approaches the experimental $M_{c}/H$ for $T > 50$ K.
However, we could not obtain $J_4$ that explained all $M/H$s along three crystal axes well.
As mentioned above, the anisotropy of the $g$-factors does not account for the anisotropy of $M/H$. 
The $J_{1}$ and $J_{2}$ interactions and the effects of AF-LRO at around $T_{\rm N}$ should be taken into account in the precise calculation.
\par

\subsection{DM interaction and WF-LRO}

\begin{figure}[h]
	\begin{center}
\includegraphics[width=.5\textwidth]{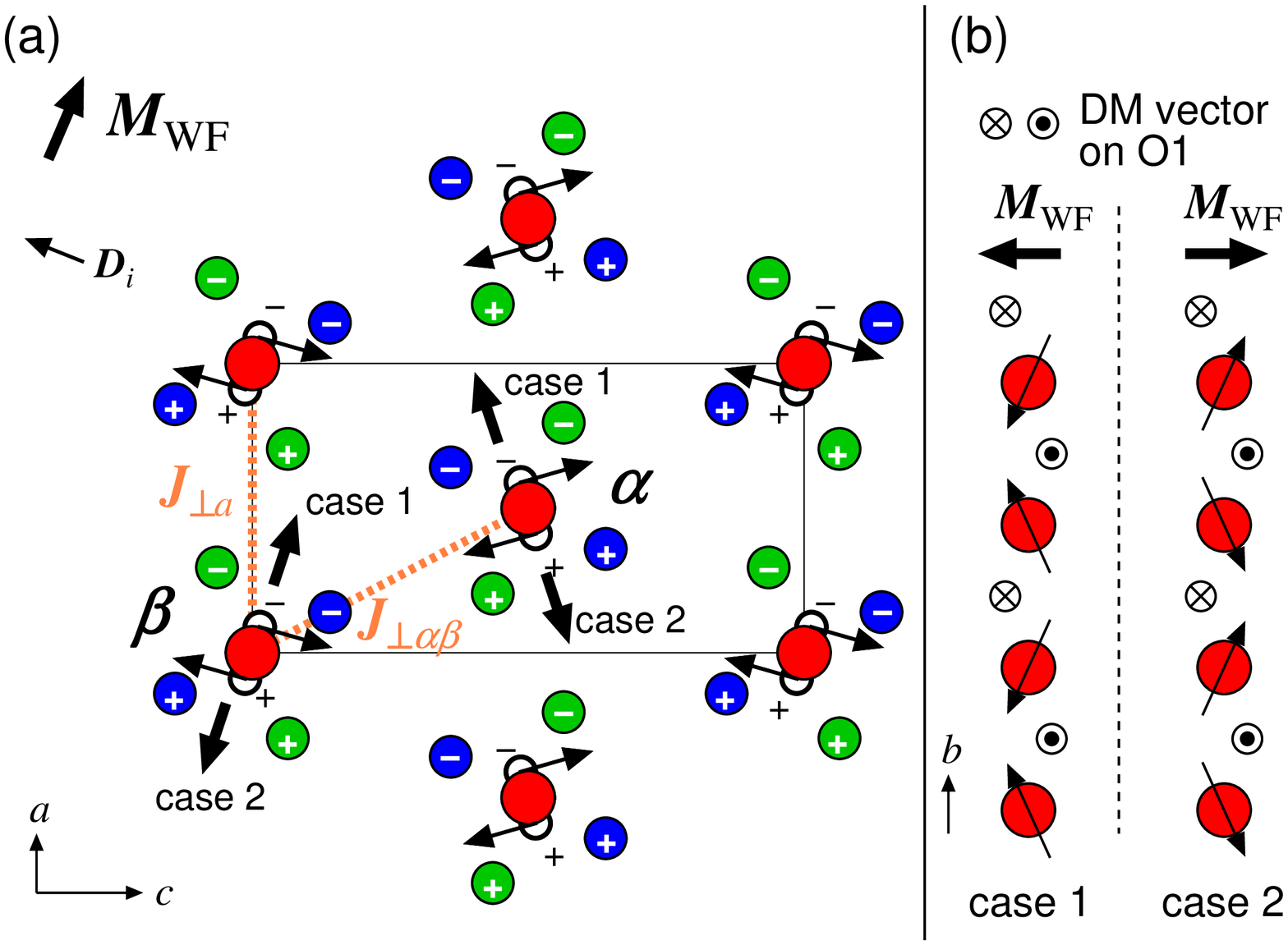}
\caption{(color online) (a)
Cu and O positions projected onto the $ac$ plane. 
Red Cu1 ions are located at $b = 0$ and $1/2$, and Cu2, Cu3, and O1 ions (marked ``+" and ``$-$") are at $b = 1/4$ and $-1/4$. 
A rectangle denotes the unit cell.
The thick arrows of \mbox{\boldmath $M$}$_{\rm WF}$ indicate the directions of the WF moments. 
The arrows indicate the direction of D vectors as shown the symbol in (b).
(b)
Two possible intrachain spin arrangements at Cu1 sites expected at $T = 0$ K. 
Symbol $\odot$ ($\otimes$) means that the D vector is directed upward (downward) with respect to the plane of the page. 
Thin arrows indicate the direction of the spin moment which is canted by the DM interaction.
}
  	\label{fig:view}
	\end{center}
  \end{figure}

Let us now consider the origin of WF-LRO.
As shown in Fig. \ref{fig:view}(a), there are two crystallographically equivalent chains. 
We refer to the tetrahedral spin chain at the center of the unit cell as the ``$\alpha$ chain" and the one at the corner as the ``$\beta$ chain". 
They have different rotation angles around the $b$ axis.

In addition to the symmetric exchange interactions $J_1$-$J_4$, there exists a DM interaction between spins at two neighboring Cu1 sites in an AF linear chain because the center of the sites is not an inversion center.
Because Cu$_{3}$Mo$_{2}$O$_{9}$ is a spin-1/2 insulating antiferromagnet, a possible origin of WF magnetism is the DM interaction.
Here we describe the Dzyaloshinskii vectors (D vectors) in bond 4 and the spatial arrangements of the vectors.
In accordance with Moriya's rule, we calculated $\bm{D}_i = D((-1)^{i}\sin \theta, 0, (-1)^{i+1}\cos \theta)$, where $\theta$ is the angle between the $c$ axis and D vector. 
We obtain $\theta = + 23.2^{\circ}$ for a Cu1-O1-Cu1 bond in the $\alpha$ chain and $ - 23.2 ^{\circ}$ for that in the $\beta$ chain, as presented in Fig. \ref{fig:view}(a), taking into account the positions of two neighboring Cu1 ions and the bridging O1 ion at room temperature. 
The factors $(-1)^i$ and $(-1)^{i+1}$ mean that two D vectors with opposite directions alternate with each other along the AF linear chain.
A chain is transformed into two neighboring ones in the $a$ direction by the translation of $\pm a$.
An $\alpha$ chain is transformed into four neighboring $\beta$ ones by the $a$-glide translation on the glide plane at $c = \pm 1/4$. 
Since the D vector is an axial one, a D vector in an $\alpha$ ($\beta$) chain is transformed into D vectors on the $ac$ plane in four neighboring $\beta$ ($\alpha$) chains, as shown in Fig. \ref{fig:view}(a). 

Let us consider the spin state below $T_{\rm N}$. 
First, we discuss the spin state of the Cu1 AF linear chain. 
The spin states of the Cu2 and Cu3 dimers are discussed later.  
We take into account the $J_4$ interaction term as the main energy and the DM interaction term in bond 4, the interchain exchange interaction terms, and the Zeeman energy as supplementary energies. 
The effects of the $J_1$ and $J_2$ interactions will be discussed later.  
In the supplementary energies, we assume that the DM interaction is the dominant one. 
As shown later, this assumption is valid. 
Because the uniform AF linear chains of Cu1 spins are formed by the $J_4$ interaction, the formation of AF-LRO in the Cu1 spins is possible with the aid of interchain interactions. 
To determine the arrangement of the spins, we have to know their principle direction.
The principle direction is defined as the spin direction when collinear AF-LRO for each chain appears at $H = 0$ T and there is no DM interaction.
The crystal symmetry is high enough so 
that the principle direction below $T_{\rm N}$ 
should be in the $a$, $b$, or $c$ direction.
Because the D vector has no $b$ component, the spin is canted toward only the $b$ direction when the principle direction is $a$ or $c$.
On the other hand, when the principle direction is $b$, the spin can be canted from the $b$ axis, which is consistent with the experimental results.
Here, we divide a magnetic moment at a Cu1 site into two components. 
The major component parallel to the chain is called the ``AF moment", and the component perpendicular to the AF moment is called the ``WF moment"(\mbox{\boldmath $M$}$_{\rm WF}$).
\par

When the DM interaction is the dominant supplementary energy, the direction of each WF moment falls into either case 1 or 2, as illustrated in Fig. \ref{fig:view}.
The directions for the two cases are exactly opposite in each chain.
Moreover, three-dimensional magnetic long-range orders successively appear at low temperatures, indicating that interchain exchange interactions cannot be ignored. 
We define two interchain exchange interaction parameters ($J_{{\bot}a}$ and $J_{{\bot}\alpha\beta}$) as represented in Fig. \ref{fig:view}(a). 
If the interchain interaction $J_{{\bot}a}$ in the $a$ direction is AF, the total magnetic moment in one chain compensates for that in the neighboring chain along the $a$ axis, and WF-LRO does not appear. 
Therefore, $J_{{\bot}a}$ is ferromagnetic, and all WF moments in $\alpha$ ($\beta$) chains fall into case 1 or 2.
Because WF-LRO appeared at $T_{\rm c} \simeq 2.5$ K at zero magnetic field in the present experiment, the interchain interaction $J_{{\bot}\alpha\beta}$ between the $\alpha$ and $\beta$ chains is ferromagnetic.
There are five patterns to explain the WF behavior of Cu$_{3}$Mo$_{2}$O$_{9}$;

{\it 1. At zero magnetic field below $T_{\rm c}$.}

In this case, both the WF-LRO and AF-LRO appear.
When all WF moments in all the chains fall into case 1 or 2, the total magnetic moment of the system has only the $a$ component, and WF-LRO parallel to the $a$ axis is realized.

{\it 2. At zero magnetic field between $T_{\rm c}$ and $T_{\rm N}$.}

In this temperature region, the jumped magnetization $\Delta M_a/g$ at zero magnetic field does not appear, but the AF-LRO and WF moment exist.
When all WF moments in all chains fall into case 1 and 2 randomly, the total magnetic moment of the system has neither the $a$ nor the $c$ component, and the WF-LRO is not realized.
We term this state a disordered state of WF moment.

{\it 3. Applying $H // a$ between $T_{\rm c}$ and $T_{\rm N}$.}

The WF-LRO was formed by applying magnetic fields.
The state of this WF moment is the same as pattern 1.
For example, at 3.0 K, the WF-LRO is stabilized when $H$ is stronger than $\sim 0.1$ T.  

{\it 4. Applying $H // c$ below $T_{\rm N}$.}

The WF-LRO was stabilized with applying $H > 0.8$ T along the $c$ axis at 2.0 K.
When all WF moments in the $\alpha$ chains fall into case 1 and those in the $\beta$ chains fall into case 2, or vice versa, the total magnetic moment of the system has only the $c$ component.
A similar phenomenon was observed in BaCu$_2$Ge$_2$O$_7$.\cite{Tsukada2000}

{\it 5. Applying $H // b$ below $T_{\rm N}$.}

Other new LRO is not realized.\\
Consequently, a WF disordered state and an AF ordered state coexist between $T_{\rm c}$ and $T_{\rm N}$ at $H = 0$ T.
\par
Let us now estimate the value of $D$ and the average absolute values of interchain exchange interactions $|J_{\bot}|$.
To estimate the value of $D$, we first have to determine the total magnetization generated by the formation of WF-LRO for $H // a$ and $H // c$ ($M^{\rm stab}_{a}$ and $M^{\rm stab}_{c}$). 
As described above, the WF-LRO at 2.0 K is completely stabilized above 0.25 and 0.8 T along the $a$ and $c$ axes, respectively, as shown in Fig. \ref{fig:expMH}.
The magnetization includes a term linearly proportional to $H$.
The linear term is derived from the magnetization below 0.3 T for $H // c$.
It cannot be estimated from $M$ in low $H$ for $H // a$ because of the rapid increase in $M$. 
Therefore, we used $M$ above 0.6 T to obtain the linear term. 
We found that $M_{a}^{\rm stab}/g = 6.1 \times 10^{-3}$ $\mu_{\rm B}$/Cu and $M_{c}^{\rm stab}/g = 2.5 \times 10^{-3}$ $\mu_{\rm B}$/Cu, as denoted by the vertical lines in Fig. \ref{fig:expMH}.
The canting angle $\phi$ of each spin from the $b$ direction was calculated from $3M_{a}^{\rm stab}/g = 0.5 \sin \phi \cos 23.2^{\circ}$ and $3M_{c}^{\rm stab}/g = 0.5 \sin \phi \sin 23.2^{\circ}$, and was estimated to be $2.36^{\circ}$ and $2.40^{\circ}$ for $H // a$ and $H // c$, respectively. 
The coarse relation $\tan 2 \phi \cong D/J_{4}$ gives $D = 20.8$ K when $J_4 \sim 250$ K. 
Here, we used the average value of $\phi$ ($2.38^{\circ}$). 
Next, we estimated $|J_{\bot}|$ by using the interchain mean-field theory.\cite{Irkhin2000}
It is determined by only $J_4$ and $T_{\rm N}$.
The result that $J_4 \sim 250$ K and $T_{\rm N} = 7.9$ K gives $|J_{\bot}| \sim 2.4$ K, which is smaller than $D$. 
Accordingly, the assumption that the DM interaction is stronger than the interchain exchange one is valid. 
Moreover, we confirmed that if $J_4$, which was not accurately determined in the present experiment, is larger than 40 K, the DM interaction is always stronger than $|J_{\bot}|$.
\par

\subsection{Spin-singlet-like spin dimer}

Let us consider the state of the Cu2 and Cu3 spins.
We refer to similar antiferromagnets.
Cu$_2$Fe$_2$Ge$_4$O$_{13}$ has spin-1/2 dimers and spin-5/2 AF uniform chains.\cite{Masuda2004}
Cu$_2$CdB$_2$O$_6$ has spin-1/2 dimers and AF uniform chains.\cite{Hase2005} 
In these antiferromagnets, spin-singlet-like pairs are formed in the dimers, while the magnetic moments in the dimers are finite even at $T = 0$ K, because they interact with the spins of the AF chains, feeling their ``alternating" internal field. 
Since the AF $J_3$ interaction is dominant and the $J_1$ and $J_2$ ones are not negligible in Cu$_3$Mo$_2$O$_9$, we infer that the Cu2 and Cu3 spins form a similar spin-singlet-like pairs. 
The finite magnetic moments of spin-singlet-like pairs are located at the Cu2 and Cu3 sites.
Taking Cu1 into account, Cu$_3$Mo$_2$O$_9$ has slightly distorted tetrahedral bonds. 
Consequently, these lead to geometrical magnetic frustration among the $J_4$, $J_1$, and $J_2$ interactions with the Cu1 spins irrespective of the signs of $J_1$ and $J_2$.
\par

\subsection{Successive phase transitions}

Finally, we discuss the mechanism of the successive phase transitions of the LROs of AF moments at $T_{\rm N}$ and of that of WF moments with respect to the arrangement of case 1 and case 2 at $T_{\rm c}$.
Similar phenomena were reported for CsNiCl$_3$ and CsNiBr$_3$, which have a triangular lattice of AF chains.\cite{Clark1972,Sano1989}
The components of the magnetic moments parallel and perpendicular to the chains are successively ordered in long range at $T_{\rm {N1}}$ and $T_{\rm {N2}}$ ($< T_{\rm {N1}}$).\cite{Clark1972,Sano1989,Maegawa1991}
Therefore, a disordered state of the perpendicular components appears between $T_{\rm {N1}}$ and $T_{\rm {N2}}$. 
This is explainable by both frustration due to the triangular lattice and anisotropy such as Ising and single-ion anisotropies.\cite{Maegawa1991,Miyashita1985} 
In Cu$_3$Mo$_2$O$_9$, there is frustration among the $J_4$, $J_1$, and $J_2$ interactions for two neighboring Cu1 spins through the Cu2 and Cu3 spins in a chain. 
The DM interaction could be the origin of the anisotropy. 
Accordingly, we infer that both the frustration and anisotropy generate AF-LRO of AF moments and disorder of WF moments between $T_{\rm N}$ and $T_{\rm c}$. 
On the other hand, the successive phase transitions are not induced without frustration.
In BaCu$_{2}$Ge$_{2}$O$_{7}$, the AF phase transition took place together with the formation of WF-LRO at $T_{\rm N} = 8.8$ K, because it has only the anisotropy of the DM interaction.\cite{Tsukada2000}

Some analogous materials, in which successive phase transitions were induced by both the frustration and DM interaction, were reported.
In Ni$_{3}$V$_{2}$O$_{8}$, two incommensurate and two commensurate phases successively appear.
The WF-LRO in the commensurate phase and ferroelectricity in the incommensurate phase have been observed.\cite{Lawes2004,Lawes2005}
CuB$_{2}$O$_{4}$ undergoes successive phase transitions to commensurate and incommensurate phases with decreasing temperature.
In the commensurate phase, the field dependence of the magnetization reveals that CuB$_{2}$O$_{4}$ is a weak ferromagnet with magnetic moments of the two antiferromagnetically coupled sublattices lying in the tetragonal plane of the crystal.\cite{Petrakovskii1999,Petrakovskii2000}
Moreover, a soliton lattice appears at around the temperature of transition to the incommensurate phase. 
The Dzyaloshinskii-Moriya interaction and anisotropy lead to the formation of a magnetic soliton lattice.\cite{Roessli2001}
These phenomena can be understood by effects of both the frustration and anisotropy.

\par

\section{CONCLUSION}
We studied the magnetic susceptibility 
$M/H$, the magnetization, and the specific heat of Cu$_3$Mo$_2$O$_9$ single crystal.
The high-temperature magnetism can be described as a sum of the isolated spin dimers and the AF linear chains in a rough approximation.
We observed an AF phase transition 
at 7.9 K together with an increase of the susceptibility along the $a$ axis.
Moreover, a jump is observed together with hysteresis in the magnetization below about 3.0 K when a tiny magnetic field along the $a$ axis is reversed. 
This result indicates that the WF phase transition occurs at $T_{\rm c} \simeq  2.5$ K at zero magnetic field, but the coercive force is very weak.
An increase of magnetization was observed along the $c$ axis with applying  magnetic field and $M_{c}$ asymptotically approached $M_{a}$.
This WF moments on the $ac$ plane can be explained by 
the Dzyaloshinskii-Moriya interaction in the Cu1 linear chains.
A component of magnetic moment parallel to the chain forms AF-LRO below $T_{\rm N}$, while a perpendicular component is disordered above $T_{\rm c}$ at zero magnetic field and forms WF-LRO below $T_{\rm c}$.
Moreover, WF-LRO is also realized with applying magnetic fields  between $T_{\rm c}$ and $T_{\rm N}$.
The separation between the AF and WF phase transitions probably comes from the frustration between the spin dimer and the AF linear chain.
\par

\section*{ACKNOWLEDGMENTS}
We are grateful to H. Yamazaki for ESR measurements, 
to M. Matsuda and K. Kakurai for neutron measurements, 
to N. Maeshima for calculation of magnetic susceptibility, 
and to H. Mamiya, T. Furubayashi, and M. Imai for use of their PPMS machine.
This work was supported by grants for basic research 
from NIMS and by a Grant-in-Aid for Scientific Research 
from the Ministry of Education, Culture, Sports, Science, 
and Technology.
\par

\end{document}